\documentclass[preprint2]{aastex63}

\usepackage{graphicx,epsfig,fancyhdr,psfig,rotating,amsmath,epsf,txfonts,natbib,epstopdf,multirow,booktabs}

\received{***}
\revised{***}
\accepted{May 7, 2021}
\submitjournal{ApJ}

\shorttitle{Formation of solar quiescent coronal loops}
\shortauthors{Hou et al.}

\def \kms {{\rm km\;s$^{-1}$}}
\def \arcsec {$^{''}$}
\def \ha {H$\alpha$}
\graphicspath{{./}}

\begin{document}
\title{Formation of solar quiescent coronal loops through magnetic reconnection in an emerging active region}

\correspondingauthor{Hui Tian}
\email{huitian@pku.edu.cn}

\author{Zhenyong Hou}
\affiliation{School of Earth and Space Sciences, Peking University, Beijing, 100871, China}

\author{Hui Tian}
\affiliation{School of Earth and Space Sciences, Peking University, Beijing, 100871, China}
\affiliation{Key Laboratory of Solar Activity, National Astronomical Observatories, Chinese Academy of Sciences, Beijing, \\
100012, China}

\author{Hechao Chen}
\affiliation{School of Earth and Space Sciences, Peking University, Beijing, 100871, China}

\author{Xiaoshuai Zhu}
\affiliation{Key Laboratory of Solar Activity, National Astronomical Observatories, Chinese Academy of Sciences, Beijing, \\
100012, China}

\author{Zhenghua Huang}
\affiliation{Shandong Provincial Key Laboratory of Optical Astronomy and Solar-Terrestrial Environment, Institute of Space Sciences, \\
Shandong University, Weihai, 264209 Shandong, People’s Republic of China}

\author{Xianyong Bai}
\affiliation{Key Laboratory of Solar Activity, National Astronomical Observatories, Chinese Academy of Sciences, Beijing, \\
100012, China}

\author{Jiansen He}
\affiliation{School of Earth and Space Sciences, Peking University, Beijing, 100871, China}

\author{Yongliang Song}
\affiliation{Key Laboratory of Solar Activity, National Astronomical Observatories, Chinese Academy of Sciences, Beijing, \\
100012, China}

\author{Lidong Xia}
\affiliation{Shandong Provincial Key Laboratory of Optical Astronomy and Solar-Terrestrial Environment, Institute of Space Sciences, \\
Shandong University, Weihai, 264209 Shandong, People’s Republic of China}

\begin{abstract}
Coronal loops are building blocks of solar active regions.
However, their formation mechanism is still not well understood.
Here we present direct observational evidence for the formation of coronal loops
  through magnetic reconnection as new magnetic fluxes emerge into the solar atmosphere.
Extreme-ultraviolet observations of the Atmospheric Imaging Assembly (AIA) onboard the Solar Dynamics Observatory (SDO)
  clearly show the newly formed loops following magnetic reconnection within a plasma sheet.
Formation of the loops is also seen in the \ha\, line-core images taken by the New Vacuum Solar Telescope.
Observations from the Helioseismic and Magnetic Imager onboard SDO show that a positive-polarity flux concentration moves towards a negative-polarity one
  with a speed of $\sim$0.4\,\kms\, before the formation of coronal loops.
During the loop formation process, we found signatures of flux cancellation and subsequent enhancement of the transverse field between the two polarities.
The three-dimensional magnetic field structure reconstructed through a magnetohydrostatic model shows field lines consistent with the loops in AIA images.
Numerous bright blobs with an average width of 1.37\,Mm appear intermittently in the plasma sheet and move upward with a projected velocity of $\sim$114\,\kms.
The temperature, emission measure and density of these blobs 
  are about 3\,MK, 2.0$\times$10$^{28}$\,cm$^{-5}$ and 1.2$\times$10$^{10}$\,cm$^{-3}$, respectively.
A power spectral analysis of these blobs indicates that the observed reconnection is likely not dominated by a turbulent process.
We have also identified flows with a velocity of 20 to 50\,\kms\,towards the footpoints of the newly formed coronal loops.
\end{abstract}
\keywords{magnetic reconnection – Sun: active region - Sun: corona}

\section{Introduction}
\label{sec:intro}

As bright structures confining heated plasma and outlining the magnetic field topology,
 plasma loops in the upper solar atmosphere have been frequently studied and they can be classified into three groups based on their temperatures, i.e. cool, warm and hot loops \citep{2014LRSP...11....4R}.
Cool loops with a temperature of 0.1--1\,MK could be observed in ultraviolet (UV) spectral lines or narrow-band images, and they have been intensively explored recently based on the Interface Region Imaging Spectrograph \citep[IRIS,][]{2014SoPh..289.2733D} observations \citep[e.g.,][]{2015ApJ...810...46H}.
Warm loops consist of plasma at a temperature of around 1--2\,MK, and they are well observed by extreme ultraviolet (EUV) imagers and spectrographs \citep[e.g.,][]{1999ApJ...517L.155L,Xie2017}.
Loops with a temperature higher than 2\,MK are defined as hot loops, which are typically observed in some spectral lines with a high formation temperature and filters with a high-temperature response, often at the wavelengths of soft X-ray, EUV and UV \citep[e.g.,][]{Winebarger2011}.
Both hot and warm loops may be called coronal loops.

Since coronal loops are building blocks of solar active regions (ARs), it is important to understand how they are formed. 
However, despite intensive investigations on the plasma properties of quiescent coronal loops, their formation process has rarely been studied and thus the formation mechanism is not well understood.
There are only a few studies on the formation of quiescent coronal loops in the literature.  
\citet{2008ApJ...679L.161M} simulated the formation of coronal loops using a class of plasma heating models. 
They found that thermal instability could lead to the dynamic formation of coronal loops that are roughly consistent with EUV observations if the plasma is sufficiently heated.
\citet{2014A&A...564A..12C} presented a model for the formation of coronal loops in an emerging AR, where the loop formation is triggered by an increase in the upward-directed Poynting flux at the loop footpoints as a result of the advection of the photospheric magnetic field.
The relationship between impulsive coronal heating in ARs and magnetic field evolution at the solar surface was investigated by \citet{2018A&A...615L...9C} and \citet{2020A&A...644A.130C}.
They found that the energy released through interactions between the opposite magnetic polarities in the photosphere could be responsible for the coronal emission in the cores of ARs.
\citet{2021ApJ...909..105T} studied the formation of some transient loops with a temperature of log ({\it T}/K) = 6.65--6.95 in the core of AR\,11890.
They conjectured that these transient loops might form through magnetic reconnection. With the help of coronal magnetic field extrapolation,   \citet{2010A&A...510A..40H} found that a magnetic reconnection process between converging magnetic fluxes lead to reconfiguration of the coronal magnetic field and enhanced soft X-ray emission in the reconnected coronal loop. 
In addition, investigating the formation process of quiescent coronal loops may help us understand various physical processes involved in flux emergence. For instance, there are suggestions that coronal loops could result from the rise of undulatory flux tubes whose dipped lower parts emerge to the corona after magnetic reconnection \citep{2004ApJ...614.1099P}.

Here we present direct observational evidence for the formation of coronal loops 
  through magnetic reconnection as new magnetic fluxes emerge to the solar atmosphere.
Several observational signatures of magnetic reconnection have been detected from multi-wavelength observations and photospheric magnetograms.
We describe our observations in Section\,\ref{sec:obs}, present the analysis results in Section\,\ref{sec:res}, 
  discuss the results in Section\,\ref{sec:dis} and summarize our findings in Section\,\ref{sec:sum}.

\section{Observations}
\label{sec:obs}

\begin{figure*}
\centering
\includegraphics[trim=0.0cm 0.8cm 0.0cm 0.5cm,width=1.0\textwidth]{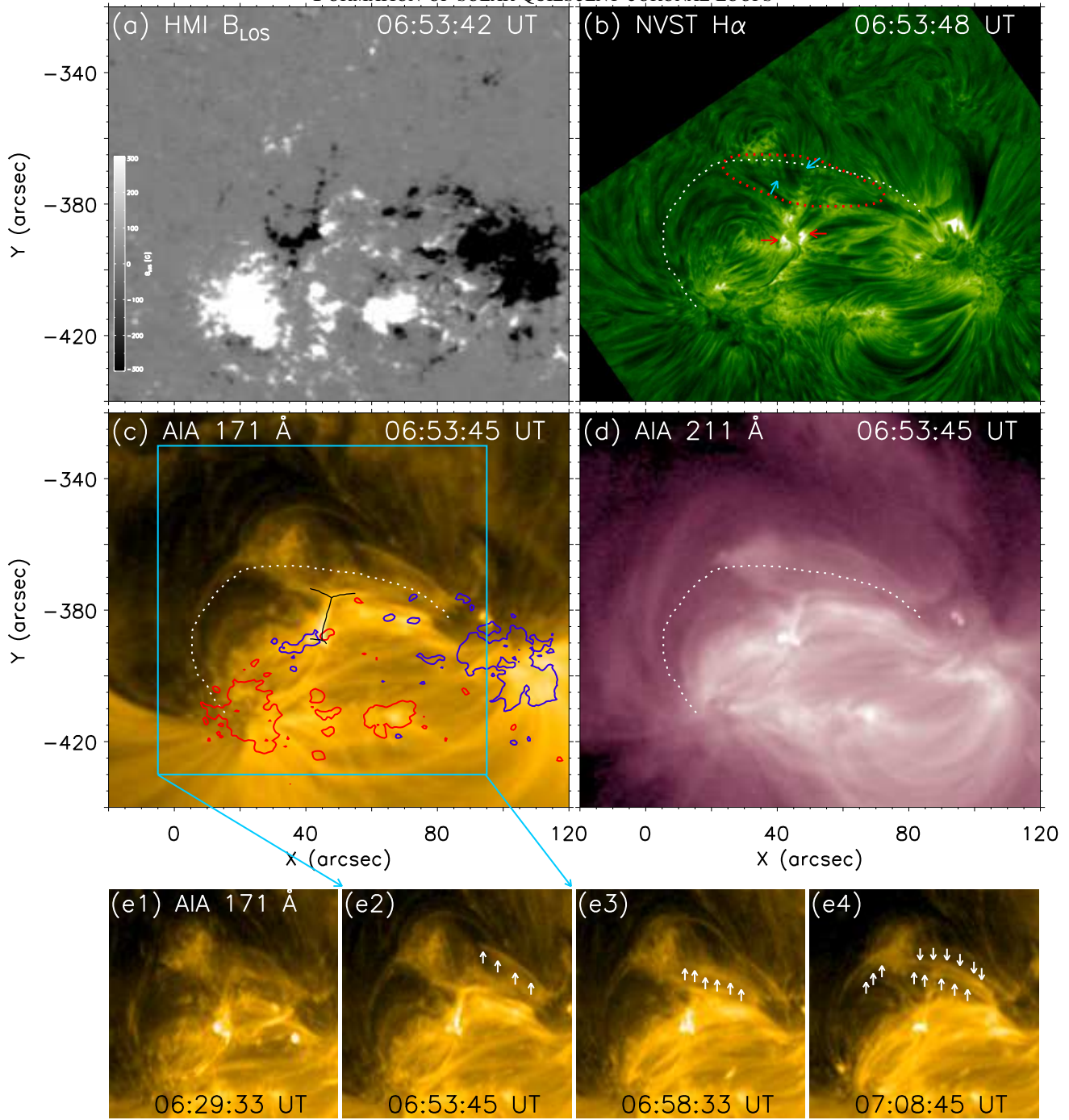}
\caption{Images of the HMI LOS magnetic field (a),
  NVST \ha\, line core (b), AIA\,171\,\AA\, (c) and AIA\,211\,\AA\, (d) taken at 06:53 UT.
  The white dotted line marks the approximate location of the newly formed overlying loops. 
The red dotted line and blue arrows in (b) indicate some dark threads that are located close to the top part of these loops.
The red arrows indicate a pair of bright ribbons that correspond to the footpoints of the newly formed small loops underneath the plasma sheet.
The red and blue contours in (c) represent positive and negative magnetic fluxes with the levels of $\pm$\,300\,G, respectively.
The black lines in (c) outline the geometry of the reconnection region.
The cyan box in (c) marks the field-of-view of the 171\,\AA\, image sequence in (e1)--(e4),
   which shows the formation process of quiescent coronal loops during the second episode.
The arrows in (e2) -- (e4) mark several newly formed coronal loops.
An animation of this figure is available, showing the formation process of coronal loops.
It includes the images of the HMI LOS magnetic field, NVST \ha\, line core, AIA\,171\,\AA\, and AIA\,211\,\AA,
  and covers 82 minutes starting at 05:49 UT with a cadence of 12\,s.}
\label{fig:overview}
\end{figure*}

The target of interest is an emerging AR with a NOAA number of 12778.
%The formation of coronal loops are observed in an emerging AR with a NOAA number of 12778 taken on 2020 October 26 from 03:00\,UT to 09:00\,UT.
The main data analyzed in this study were taken from 03:00\,UT to 09:00\,UT on 2020 October 26, by the Atmospheric Imaging Assembly \cite[AIA,][]{2012SoPh..275...17L} and 
  the Helioseismic and Magnetic Imager \citep[HMI,][]{Schou2012hmi} onboard the Solar Dynamics Observatory \citep[SDO,][]{2012SoPh..275....3P}.
We used the images taken in the UV channel of 1600\,\AA\, and 
  the EUV channels of 171\,\AA, 193\,\AA, 211\,\AA, 131\,\AA, 335\,\AA\, and 94\,\AA\,of AIA to study the formation of coronal loops and the associated reconnection process.
The pixel size of the AIA images is 0.6\arcsec, while the cadences are 24\,s for the UV channel and 12\,s for the EUV channels.
To investigate the evolution of the magnetic field topology during the reconnection process, we used the line-of-sight (LOS) and the full-disk disambiguated vector magnetic field data \citep[‘hmi.B 720s’ data series,][]{2014SoPh..289.3483H} taken by HMI.
The pixel size of the magnetograms is 0.5\arcsec, and the cadences are 45\,s and 720\,s for the LOS magnetograms and the vector magnetograms, respectively.
The Very Fast Inversion of Stokes Vector\,\citep{ 2011SoPh..273..267B} has been performed for the full-disk vector magnetic field.
After that, the azimuth angle of the magnetic field is determined by a minimum energy method \citep{1994SoPh..155..235M, 2009SoPh..260...83L}.
%The vector magnetic field has been obtained through the Very Fast Inversion of Stokes Vector\,\citep{ 2011SoPh..273..267B}, which is a Milne-Eddington based algorithm.
%A minimum energy method\,\citep{1994SoPh..155..235M, 2009SoPh..260...83L} has beed applied to resolve the 180$^{\circ}$ ambiguity in the transverse field.

Chromospheric images of \ha\, line core, taken by the ground-based New Vacuum Solar Telescope \citep[NVST,][]{Liu_2014,Yan2020} from 03:25\,UT to 08:58\,UT on 2020 October 26, were also used to examine dynamics of the relatively cool plasma during the loop formation process.
These images have a cadence of $\sim$10\,s and a pixel size of 0.165\arcsec. Standard calibration and reconstruction for the NVST images have been applied \citep{2016NewA...49....8X}.

\section{Results}
\label{sec:res}

Figure\,\ref{fig:overview} presents an overview of the observed AR at 06:53\,UT  during its emerging phase.
The LOS magnetogram shown in Fig.\,\ref{fig:overview}\,(a) reveals two large flux concentrations with opposite polarities and numerous smaller ones between them.
Many fine filamentary structures connecting these magnetic fluxes can be seen from the \ha\, line core image (Fig.\,\ref{fig:overview}\,(b)).
The coronal emission is dominated by bright plasma loops, as observed in the channels of AIA\,171\,\AA\, and 211\,\AA.

\begin{figure*}
\centering
\includegraphics[trim=0.0cm 0.0cm 0.0cm 0.0cm,width=0.9\textwidth]{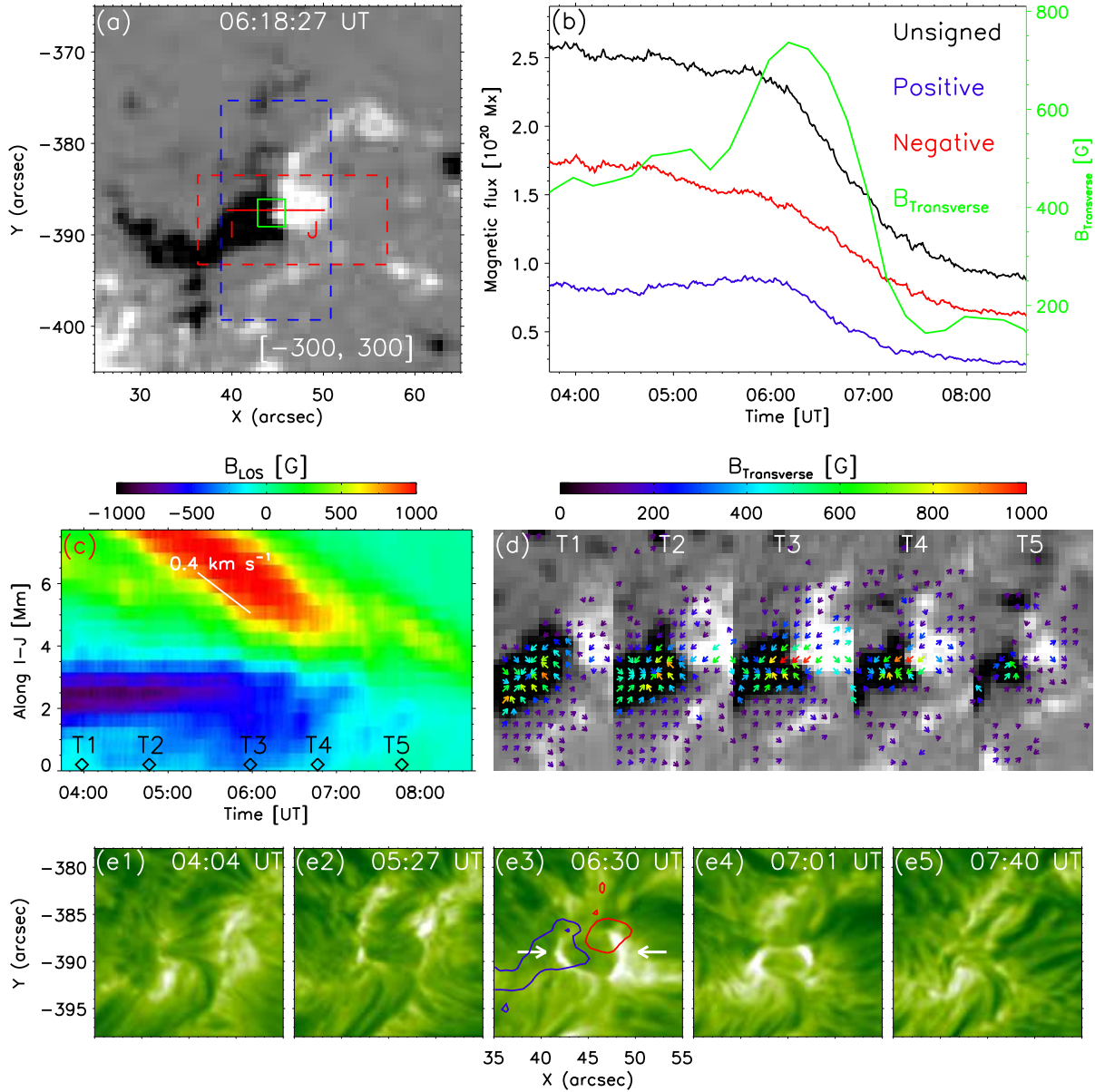}
\caption{Evolution of the photospheric magnetic field.
(a) LOS magnetogram observed at 06:18:27\,UT, saturated at $\pm$\,300\,G.
(b) Temporal evolution of the positive (blue), negative (red) and unsigned (black) magnetic fluxes within the red rectangle in (a) as well as the averaged transverse field strength (green) within the green rectangle in (a).
(c) Time-distance image of the LOS magnetic field for the slice (I--J) marked by the red line in (a).
The white solid line indicates the trajectory of the positive polarity, and the corresponding speed is also shown.
(d) Sequence of the photospheric vector magnetogram for the region of the blue rectangle in (a).
The observation times (T1--T5) of the five magnetograms are indicated by the diamonds in (c).
The background images show the magnitude of the radial component (B${_r}$) of the field.
The directions and colors of the arrows indicate the directions and magnitudes of the transverse magnetic field ($\geqslant$100\,G), respectively. 
(e1)--(e5) Sequence of the \ha\, line core images.
The arrows indicate the pair of bright ribbons that correspond to the footpoints of the newly formed small loops.
The red and blue contours represent positive and negative magnetic fluxes with the levels of $\pm$\,300\,G, respectively.
}
\label{fig:mg}
\end{figure*}

\subsection{Formation of coronal loops}
\label{subsec:formation}

From the associated animation of Fig.\,\ref{fig:overview}, we can see a bundle of coronal loops form during the period of 06:00\,UT to 07:10\,UT.
Two episodes of loop formation can be seen during 06:00\,UT$-$06:19\,UT and 06:41\,UT$-$07:10\,UT, respectively.
The image sequence presented in Fig.\,\ref{fig:overview}(e1)--(e4) show the formation of coronal loops during the second episode.
These newly formed loops are clearly visible in the AIA\,304\,\AA, 171\,\AA, 193\,\AA\, and 211\,\AA\, passbands, and are less obvious in 335\,\AA\, and 131\,\AA.
Due to the dynamic nature and low contrast with respect to the strong background coronal emission, these loops cannot be easily identified from still images.
So in Fig.\,\ref{fig:overview} we used a white dotted curve to indicate the approximate location of these loops. 
A bright lane, presumably a plasma sheet, appears below the newly formed coronal loops.
Numerous bright plasma blobs were found to move upward along the plasma sheet during the two episodes.
The plasma sheet can be clearly identified from all AIA EUV channels, especially the 304\,\AA, 171\,\AA, 193\,\AA\, and 211\,\AA\,channels,
while no counterpart appears in the AIA\,1600\,\AA\, passband.

Above (i.e. to the north) the plasma sheet, several segments of enhanced EUV emission appear as the plasma blobs move upward, delineating the top part of the newly formed coronal loops.
The loop tops appear as dark threads in the \ha\, core images, and are indicated by the blue arrows in Fig.\,\ref{fig:overview}\,(b).
These segments of plasma clearly move downwards to the legs and footpoints of the newly formed coronal loops.
In addition to these large, newly formed coronal loops, an ensemble of bright, small loop-like structures also appear simultaneously below the plasma sheet, as seen from the AIA EUV images.
In the meantime, the \ha\, core images reveal some dark fibrils connecting two bright ribbons, as denoted by the two red arrows in Fig.\,\ref{fig:overview}\,(b). 
On some occasions, a cusp-like structure can be seen at one or two ends of the plasma sheet (Fig.\,\ref{fig:overview}\,(c)\&(d)).

The footpoints of the newly formed coronal loops are located in or around the two large flux concentrations,
  while the small, newly formed loops are rooted in a region with opposite magnetic polarities around the location of solar\_x\,=\,45\arcsec\, and solar\_y\,=\,-387\arcsec.
Before and during the formation process of the overlying coronal loops, another two groups of loops/fibrils (see in AIA\,171, 211\,\AA\, and \ha\, core, side loops thereafter)
  exist to the east and west of the plasma sheet.
The dynamic evolution of the loop configuration, the plasma sheet, the moving plasma blobs and the opposite magnetic polarities mentioned above suggest that 
  the overlying coronal loops are likely formed through magnetic reconnection between the two side loops.

\begin{figure*}[!ht]
\centering
\includegraphics[trim=0.0cm 0.0cm 0.0cm 0.0cm,width=0.8\textwidth]{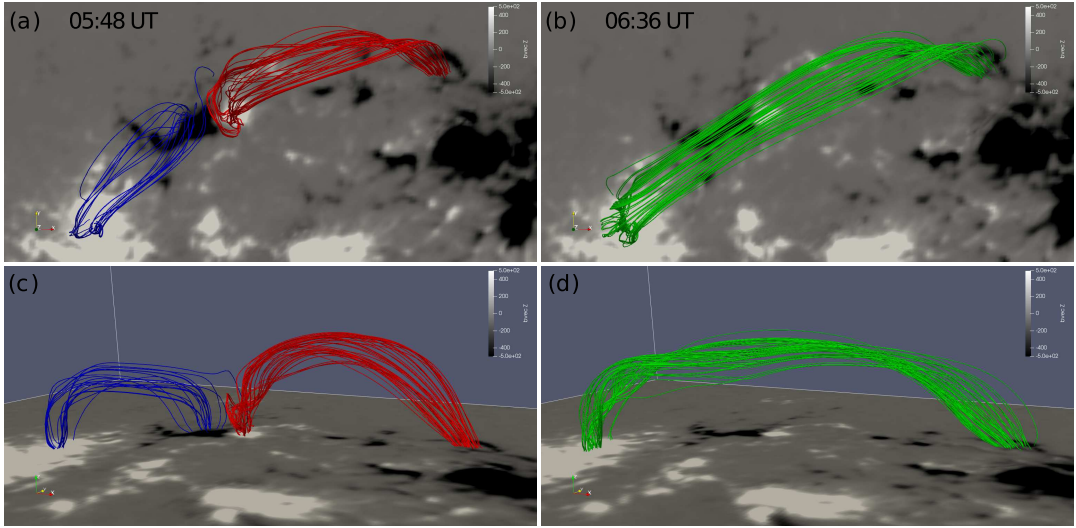}
\caption{Magnetic field lines in the MHS model. Results based on photospheric magnetograms taken at 05:48\,UT and 06:36\,UT are shown in the left and right panels, respectively.
The top and side views are presented in the top and bottom rows, respectively.
The red/blue field lines in the left panels and the green lines in the right panels are traced from the same positions.
The background images show the HMI LOS magnetograms.
}
\label{fig:flines}
\end{figure*}

\subsection{Magnetic field evolution and geometry of magnetic reconnection}
\label{subsec:mgrec}

To further confirm that the coronal loops are formed through magnetic reconnection, we have analyzed the magnetic field data in detail.
From the HMI LOS magnetograms, we can see that AR\,12778 appears first around 01:00\,UT on 2020 October 24. Roughly two days later, two large flux concentrations form. At this time the magnetic fluxes within the AR still keep emerging and the two flux concentrations also keep growing.
During the process of flux emergence, two small patches with opposite magnetic polarities tend to move towards each other and 
  meet around the location of solar\_x\,=\,45\arcsec\, and solar\_y\,=\,-387\arcsec\, (see Fig.\,\ref{fig:mg}\,(a)) at about 04:00\,UT on 2020 October 26.
Before 2020 October 26, the negative and positive polarities move with a speed of 0.2\,\kms\,and 0.4\,\kms, respectively.

The two approaching magnetic patches are highlighted by the red rectangle in Fig.\,\ref{fig:mg}(a). Before 06:00\,UT on 2020 October 26, the magnetic fluxes of these two patches are more or less stable. However, an obvious signature of flux cancellation between the two patches can be seen from the HMI LOS magnetograms (see the associated animation of Fig.\,\ref{fig:overview}) after 06:00\,UT, suggesting the occurrence of magnetic reconnection in the low atmosphere \citep[e.g.,][]{1993SoPh..143..119W,2018ApJ...854..174T,2018ApJ...861..135Y}. Fig.\,\ref{fig:mg}\,(b) shows a similar decreasing trend in the positive, negative and 
unsigned magnetic fluxes. During the period of 06:00\,UT-07:10\,UT, when the formation of coronal loops is observed, the unsigned magnetic flux decreases from 2.3$\times$$10^{20}$ Mx to 1.2$\times$$10^{20}$ Mx, with an average cancellation rate of $\sim$$10^{20}$ Mx hr$^{-1}$. 

We put one slice\,I-J (marked by the red horizontal line in Fig.\,\ref{fig:mg}\,(a)) across the two opposite-polarity magnetic patches, and present the time-distance image for the slice in Fig.\,\ref{fig:mg}\,(c).
During the period of $\sim$05:00\,UT to 06:00\,UT, the negative-polarity patch is nearly stationary, while the positive one moves further towards the negative flux with a speed of $\sim$0.4\,\kms.
Fig.\,\ref{fig:mg}\,(d) shows the temporal evolution of the vector magnetic field. After 06:00 UT, flux cancellation is clearly seen from the LOS magnetograms. 
The transverse field around the interface of the two polarities is weak and shows different directions at the time T1.
Starting from the time T2, a transverse field pointing from the positive polarity to the negative one is built.
As new coronal loops are sequentially formed, i.e.,  at times T3 and T4, the transverse field is enhanced by $\sim$260\,G.
At time T5, the formation of new coronal loops ceases.
As a result, the transverse field becomes weaker and there is no uniform direction again.
The green curve in Fig.\,\ref{fig:mg}\,(b) clearly shows this trend of enhanced and then reduced transverse field.

Our observations of the converging opposite magnetic polarities, the subsequent flux cancellation and simultaneous enhancement of the transverse field
  between the two polarities indicate that the coronal loops are formed through magnetic reconnection.
No sign of the plasma sheet is present in the AIA\,UV channels, indicates that the reconnection might occur in the corona.
As mentioned in section\,\ref{subsec:formation} and shown in Fig.\,\ref{fig:mg}\,(e2)--(e4), the chromospheric \ha\, core images show some dark,
  thin fibrils below the plasma sheet and connecting the two bright ribbons. 
After the formation of coronal loops, these dark fibrils and bright ribbons disappear around 07:40\,UT (see Fig.\,\ref{fig:mg}\,(e5)), indicating submergence of the small loops.

%This model extrapolates the magnetic field by a magnetohydrodynamic relaxation approach.
%It is particularly appropriate for layers where the plasma is relatively high and the force-free assumption fails such as in our case.
%copied from Song 2020
We also reconstructed the three-dimensional magnetic field topology through a magnetohydrostatic (MHS) model developped by \citet{2018ApJ...866..130Z,2019A&A...631A.162Z}.
The vector magnetograms taken by HMI at 05:48\,UT and 06:36\,UT (before and after the first episode of reconnection) were used as inputs of the model.
The magnetic field lines at the two different times are shown in Fig.\,\ref{fig:flines}.
A comparison between field lines at these two instances suggests that magnetic reconnection between the red and blue field lines lead to the formation of the green field lines.
The red/blue lines and green lines correspond to the side loops and newly formed coronal loops in AIA EUV images, respectively. 

\subsection{Plasma flows resulting from the magnetic reconnection}
\label{subsec:details}

\begin{figure*}
\centering
\includegraphics[trim=0.0cm 0.5cm 0.0cm 0.0cm,width=0.7\textwidth]{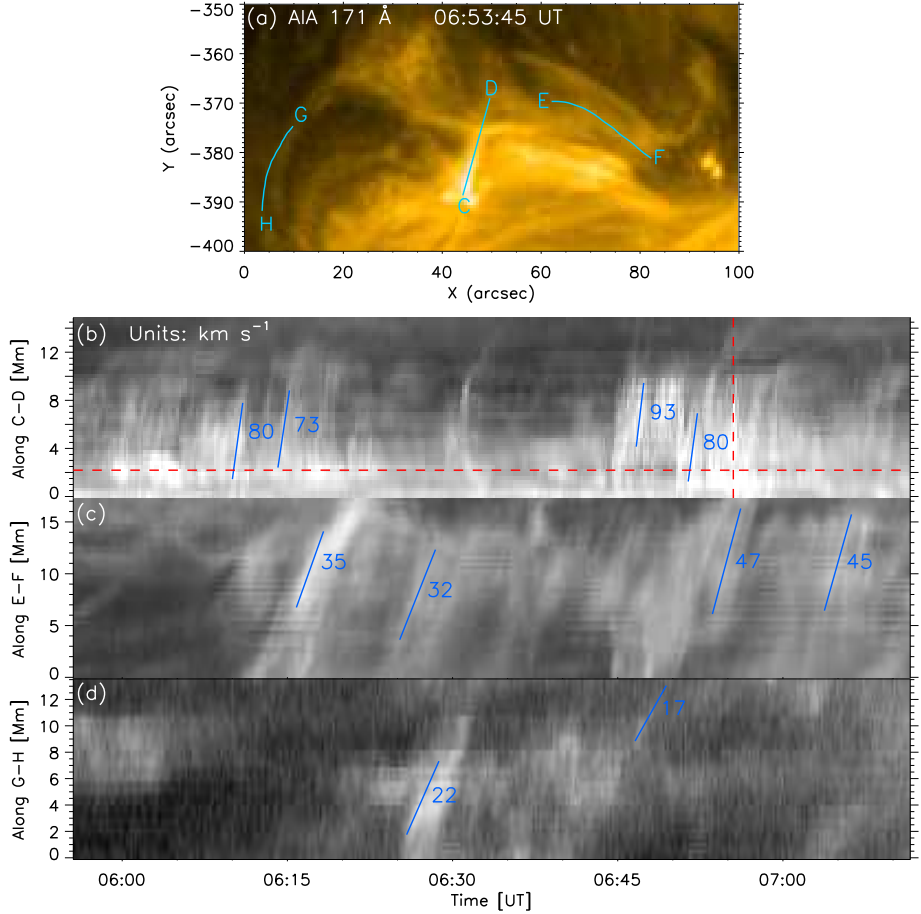}
\caption{Flows resulting from the reconnection.
(a) An AIA\,171\,\AA\, image taken at 06:53:45\,UT.
Three cyan lines C--D, E--F and G--H indicate locations of the plasma sheet and the two legs of the newly formed coronal loops, respectively. 
The time-distance images of AIA\,171\,\AA\, for slices C--D, E--F and G--H are shown in (b), (c) and (d), respectively.
The blue lines mark the tracks of plasma motions, and the numbers are the projected speeds of the flows in the unit of \,\kms.
The two red dashed lines in \,(c) indicate the time and location corresponding to Fig.\,\ref{fig:power_index}(a) and (c), respectively.
}
\label{fig:slices_stimg}
\end{figure*}

Plasma flows resulting from the magnetic reconnection can be seen in most EUV channels of AIA and \ha\, line core (see the associated animation of Fig.\,\ref{fig:overview}). Discrete bright blobs generated by the reconnection move upward along the plasma sheet. In the meantime, obvious plasma flows move downward from the top part of the newly formed coronal loops to their legs and footpoints .
We put three slices (marked by the cyan lines C--D, E--F and G--H in Fig\,\ref{fig:slices_stimg}\,(a)) at the locations of the plasma sheet and loop legs to track  the flows observed in the AIA\,171\,\AA\,channel.
The time-distance image for slice C--D is shown in Fig\,\ref{fig:slices_stimg}\,(b), where we can see numerous tracks of blobs in the plasma sheet.
As an example, the projected speeds of four blobs were estimated as the slopes of the four blue lines marked in Fig\,\ref{fig:slices_stimg}\,(b), which are 80, 73, 93 and 80\,\kms, respectively.
From this time-distance image, we can also see that the length of the plasma sheet varies with time.
The tracks of plasma flows along the legs of the newly formed coronal loops are shown in Fig\,\ref{fig:slices_stimg}\,(c)--(d).
The flows move downward along the western legs with a speed of $\sim$40\,\kms,
  while fewer flows are seen to move along the eastern legs with a speed of less than 30\,\kms.

\begin{figure*}
\centering
\includegraphics[trim=0.0cm 0.5cm 0.0cm 0.0cm,width=0.7\textwidth]{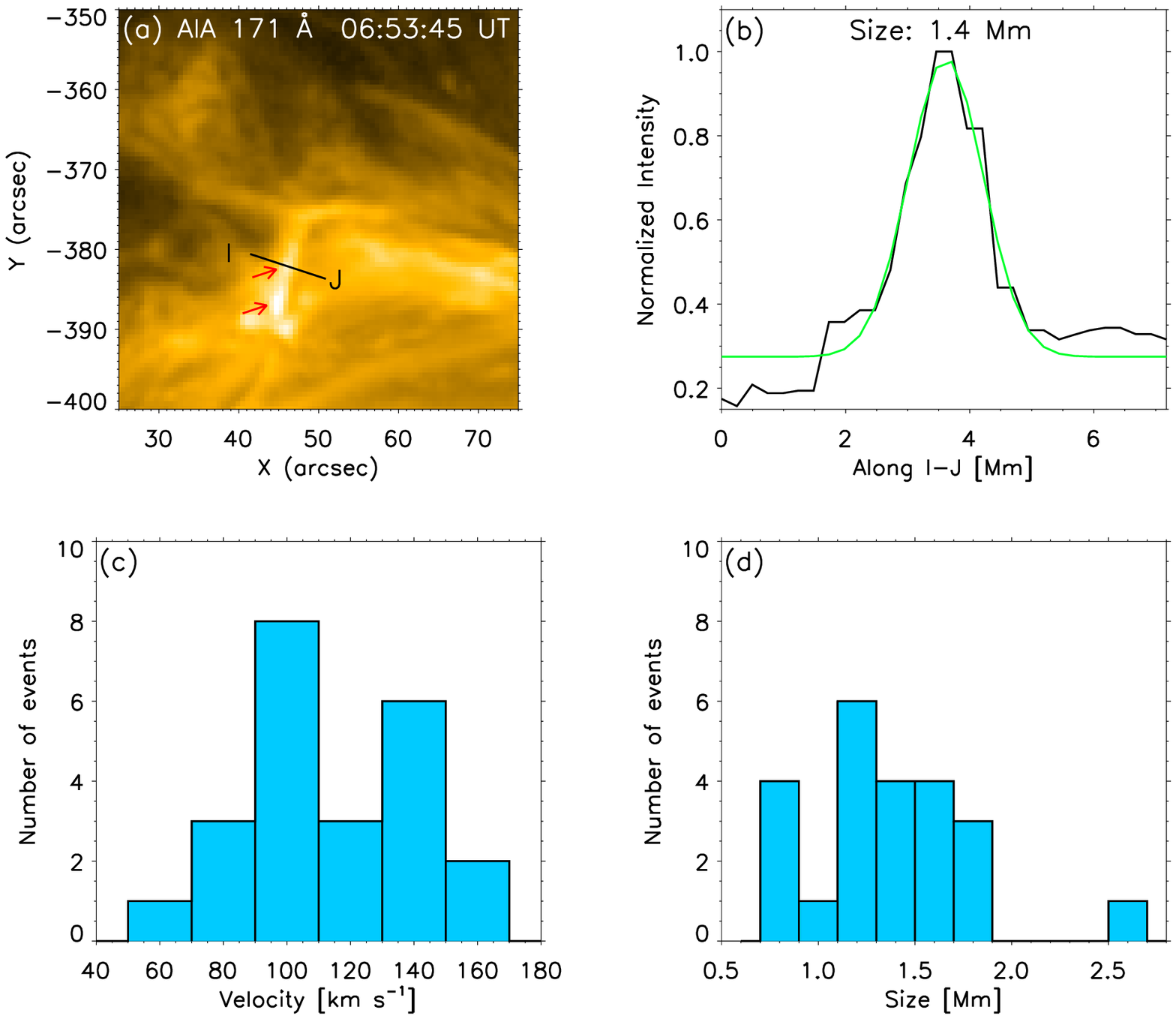}
\caption{Velocities and sizes of plasma blobs in the plasma sheet.
(a) An AIA\,171\,\AA\, image taken at 06:53:45\,UT.
(b) AIA\,171\,\AA\,intensity along slice I--J shown in (a). The black and green curves indicate the original intensity profile and the Gaussian fit.
(c) and (d) Distributions of the projected velocity and size for the identified plasma blobs.}
\label{fig:blob}
\end{figure*}

We have identified 23 well isolated blobs in the plasma sheet from the AIA\,171\,\AA\, images.
Two examples are indicated by the red arrows in Fig.\,\ref{fig:blob}\,(a).
To derive the size of each blob, we put a line that is perpendicular to the plasma sheet across the blob, e.g., the black line I--J in Fig\,\ref{fig:blob}\,(a),  and plot the AIA\,171\,\AA\,intensity along the line. 
Then we applied a single-Gaussian fit to the intensity profile, and the full-width-at-half-maximum (FWHM) was taken as the blob size. 
Fig\,\ref{fig:blob}\,(b) shows one example of the intensity profile and the fitting result.
The distributions of the size and projected velocity for these identified blobs are shown in Fig\,\ref{fig:blob}\,(c) and (d).
The blob velocity is mostly in the range of 60 to 160\,\kms, with an average of 114\,\kms. The blob size is 1.37$\pm$0.42\,Mm.

\begin{figure*}
\centering
\includegraphics[trim=0.0cm 0.0cm 0.0cm 0.0cm,width=0.7\textwidth]{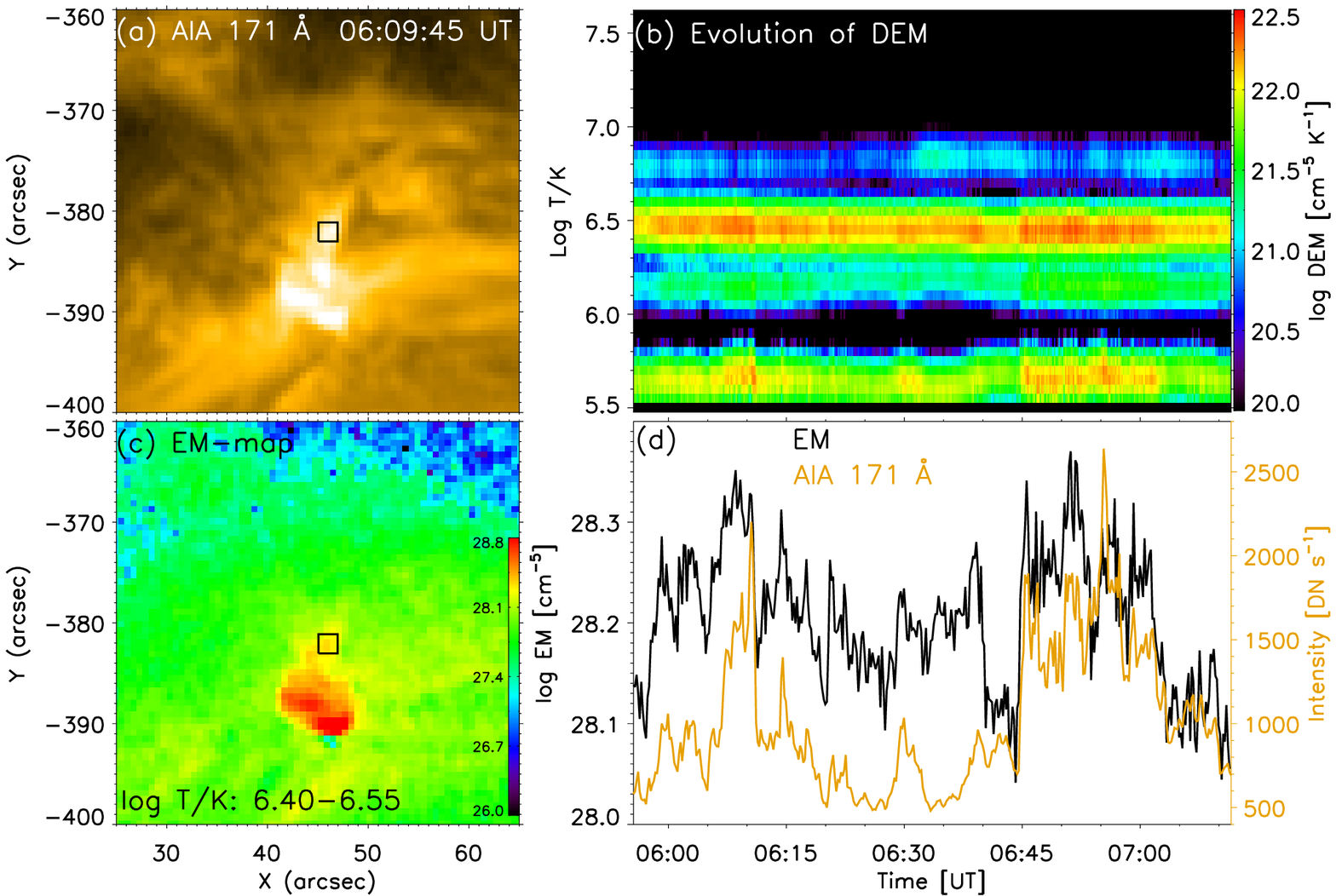}
\caption{DEM analysis for the reconnection region.
(a) An AIA\,171\,\AA\, image taken at 06:09:45\,UT.
(b) Evolution of the DEM averaged over the region outlined by the black box in (a).
(c) Map of total EM in the temperature range of log ({\it T}/K) = 6.40$-$6.55.
(d) Light curves of AIA\,171\,\AA\,intensity (yellow) and total EM (black) averaged over the region outlined by the black box in (a).
}
\label{fig:dem}
\end{figure*}

With observations in the AIA\,94\,\AA, 131\,\AA, 171\,\AA, 193\,\AA, 211\,\AA\, and 335\,\AA\,channels,
  we performed a differential emission measure \citep[DEM, see][]{2015ApJ...807..143C,2018ApJ...856L..17S,2020ApJ...898...88X,2021Innov...200083S} analysis to investigate plasma properties in the reconnection region.
The results are shown in Fig.\,\ref{fig:dem}.
Similar to \citet{2020ApJ...901...64K}, we show the temporal evolution of the DEM of the plasma sheet in Fig.\,\ref{fig:dem}\,(b).
The DEM of the plasma sheet appears to peak in the temperature range of log ({\it T}/K) = 6.40$-$6.55.
The peak temperature shows no obvious variation from 05:55\,UT to 07:12\,UT (Fig.\,\ref{fig:dem}\,(b)).
By integrating a DEM curve over the temperature range of log ({\it T}/K) = 6.40$-$6.55, we obtained a total emission measure (EM).
The spatial distribution of the total EM is presented in Fig.\,\ref{fig:dem}\,(c), where we can see enhanced EM in the reconnection region.
A similar temporal variation was found in the light curves of AIA\,171\,\AA\,intensity and total EM in the plasma sheet (Fig.\,\ref{fig:dem}\,(d)).
Peaks in the total EM curve generally correspond to the passage of plasma blobs. The blobs have a typical EM of 2.0$\times$10$^{28}$ cm$^{-5}$.
Taking the average size of the blob (1.37\,Mm) as the LOS integration length, the electron density was estimated to be 1.2$\times$10$^{10}$ cm$^{-3}$.

\begin{figure*}
\centering
\includegraphics[trim=0.0cm 0.3cm 0.0cm 0.0cm,width=0.6\textwidth]{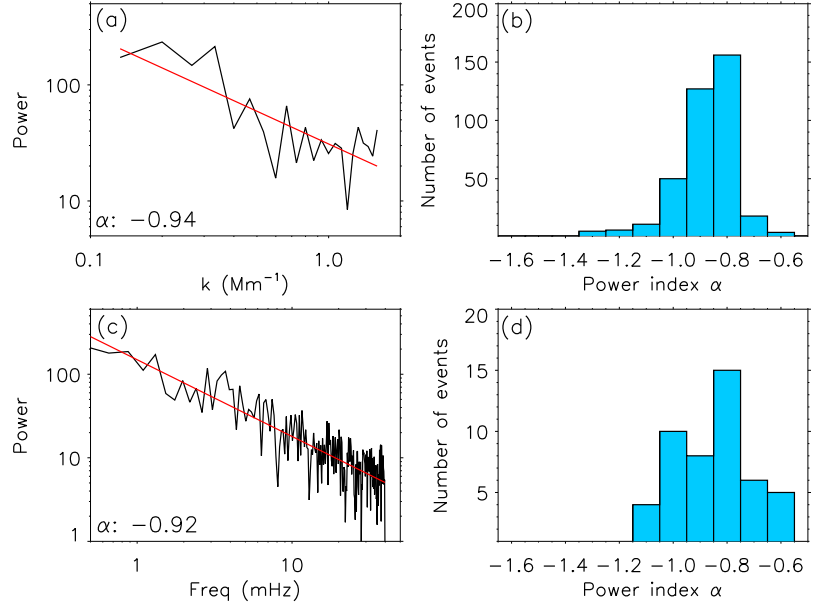}
\caption{Power spectral analysis for the plasma sheet.
(a) Power spectrum of the intensity profile along the red vertical line in Fig.\,\ref{fig:slices_stimg}\,(c) in the spatial frequency domain.
The fitting spectral index $\alpha$ is -0.94 in the range of 0.13$-$1.60\,Mm$^{-1}$.
(b) Distribution of $\alpha$ for the power spectra in spatial frequency domain.
(c) Same as (a) but for the time series along the red horizontal line in Fig.\,\ref{fig:slices_stimg}\,(c) and in the temporal frequency domain.
The fitting spectral index $\alpha$ in the frequency range of 0.44$-$39.81\,mHz is -0.92.
(d) Distribution of $\alpha$ for the power spectra in temporal frequency domain.
}
\label{fig:power_index}
\end{figure*}

We also applied the fast Fourier transform (FFT) analysis to the AIA\,171\,\AA\,emission in the plasma sheet.
As an example, Fig.\,\ref{fig:power_index}\,(a) presents a power spectrum of the intensity profile in the spatial frequency domain after FFT.
It reveals a power-law distribution with a spectral index ($\alpha$) of -0.94.
We calculated $\alpha$ for the same position in every image taken from 05:55\,UT to 07:12\,UT, and present the distribution of $\alpha$ in Fig.\,\ref{fig:power_index}\,(a).
We also applied FFT to the time series of AIA\,171\,\AA\,intensity at different positions along the plasma sheet.
One example power spectrum and the distribution of spectral index are shown in Fig.\,\ref{fig:power_index}\,(c) and Fig.\,\ref{fig:power_index}\,(d), respectively.
We found that the $\alpha$ values are mostly larger than -1.1 and distinctly different from -1.67,
  a spectral index expected in the scenario of turbulent magnetic reconnection \citep{2011ApJ...737...24B,2011ApJ...737...14S}.

\section{Discussion}
\label{sec:dis}

A combination of the magnetic field extrapolation and EUV observations suggests that the plasma sheet is located between two sets of side loops.
Since the magnetic field lines have different polarities between the two approaching side loops, electric current would be built at their interface.
This suggests that the bright plasma sheet is likely a current sheet or contains a current sheet,
  which can also be observed at different spatial scales \citep[e.g.,][]{2014ApJ...797L..14T,2015ApJ...798L..11Y,2018ApJ...858L...4X,2018ApJ...853L..18Y,2018ApJ...853L..15L}.
The thickness of the plasma sheet is $\sim$1.4 Mm, which is comparable to the thickness of the plasma sheet in a small C1.6 flare \citep{2014ApJ...797L..14T}
  and in two chromospheric reconnection events \citep{2015ApJ...798L..11Y,2018ApJ...858L...4X}.
It is much smaller than the typical thickness of current sheets measured at a distance of $\sim$1.5 solar radii or beyond during coronal mass ejections \citep[CMEs, e.g.,][]{Lin2009}.
This thickness is apparently much larger than the proton gyroradius (orders of 100 meters) in the corona, the expected reconnection thickness of a current sheet based on laboratory measurements.
It was suggested that plasmoid instability may play a critical role in broadening the CME current sheets \citep[e.g.,][]{Lin2009}.
The blob-like features observed here are likely manifestations of plasmoids, and thus the large thickness of the plasma sheet or current sheet might be caused by a similar process.
Such blobs are frequently seen in numerical simulations of magnetic reconnection involving plasmoid instability in the non-flaring solar atmosphere \citep[e.g.,][]{2015ApJ...799...79N,2020RSPSA.47690867N,2021A&A...646A..88N,2019A&A...628A...8P}.
We noticed that similar moving blob-like features within current sheets or along newly formed loops have been previously reported from observations of flares
  \citep[e.g.,][]{2012ApJ...745L...6T,2016ApJ...828..103T,2014ApJ...797L..14T,2018ApJ...866...64C,2018ApJ...853L..26H,2019SciA....5.7004G,2020SoPh..295..167M},
  filament eruptions \citep[e.g.,][]{2016NatPh..12..847L,2019ApJ...879...74C,2020A&A...633A.121X}, chromospheric jets \citep[e.g.,][]{2015ApJ...804...69Y} and coronal jets \citep[e.g.,][]{2019ApJ...870..113Z}.
To the best of our knowledge, this is the first time that plasma blobs are identified during the formation of quiescent coronal loops.
The sizes and velocities of our identified plasma blobs are of the same order of those in these flare and jet observations. 

The power spectral analysis reveals a spectral index that is obviously different from -1.67, a value found in a large-scale CME current sheet \citep{2018ApJ...866...64C}.
\citet{2018SoPh..293....6F} found that the spectral index of velocity fluctuation can change from larger than -0.5 to about -1.5 during the occurrence of CMEs, implying a rapid development of turbulence after the CME onset. Our observed spectral indexes of about -1 indicate that the fluctuations fall in the energy containing range outside the inertial range with cascading process in the traditional turbulence scenario \citep{2019LRSP...16....5V}. The observed power-law spectrum ($\sim$f$^{-1}$)  suggests that the occurrence of reconnection events is randomly intermittent instead of driven by the turbulence with nonlinear interactions between vortices \citep{2020ApJ...901L..22Y,2021ApJ...908..237H}.
Assuming that the reconnection inflow speed can be approximated as half of the converging speed and taking a typical speed of the plasma blobs as the outflow speed, we can roughly estimate the reconnection rate as 0.2/100=0.002. This reconnection rate is one to two orders of magnitude lower than those inferred from many other reconnection observations in the solar atmosphere 
  \citep[e.g.,][]{2013NatPh...9..489S,2014ApJ...797L..14T,2016NatCo...711837X}.
However, it still lies at the lower end of the possible range of reconnection rate estimated by \citet{2015NatCo...6.7598S}.

The approaching opposite-polarity magnetic fluxes and subsequent flux cancellation, the formation of overlying coronal loops above the plasma sheet, the enhancement of the transverse magnetic field, together with the submergence/disappearance of the small low-lying loops/fibrils beneath the sheet suggest a typical magnetic reconnection process between two sets of magnetic loops.
Considering the fact that the observed process occurs in an emerging AR,
  it is very likely that the two reconnecting loops correspond to two segments of an undulatory magnetic flux tube rooted in the two large flux concentrations in AR\,12778.
The middle part of the undulatory tube is likely dipped under the photosphere, and thus has a U-shape. Such a magnetic field structure is called a bald patch.
The intersections between the U-shape field lines and the photosphere correspond to the two small converging opposite-polarity magnetic patches.
When the two patches are pushed together, a current sheet develops at their interface and magnetic reconnection occurs.
In this case, large-scale overlying magnetic loops and small low-lying loops are produced, which appear as the rising, newly formed coronal loops and submerging small loops/fibrils, respectively.

Such a process is similar to the scenario suggested by \citet{2004ApJ...614.1099P}. They performed a topological analysis for the emerging AR\,8844, and found that many Ellerman bombs \citep[EBs,][]{1917ApJ....46..298E} and bald patches were connected by a series of flux tubes.
Based on this finding, they suggested that coronal loops could result from the rise of undulatory flux tubes whose dipped lower parts emerge to the corona after magnetic reconnection.
With observations of the Interface Region Imaging Spectragraph \citep[IRIS,][]{2014SoPh..289.2733D}, \citet{2014Sci...346C.315P} identified four UV bursts (UBs) and suggested that they might be produced by magnetic reconnection in U-shape structures during flux emergence in the lower solar atmosphere.
Subsequent studies revealed that these UBs indeed mostly occur in emerging flux regions \citep[e.g.,][]{2015ApJ...812...11V,2016ApJ...824...96T,2018ApJ...854...92T,2016MNRAS.463.2190N,2018SSRv..214..120Y,2019ApJ...875L..30C} and that some of them are indeed associated with U-shape magnetic field structures \citep[e.g.,][]{2016A&A...593A..32G,2017ApJ...836...52Z,2018ApJ...854..174T,2019ScChE..62.1555C,2020ApJ...893L..13S}.
The difference is that EBs and UBs are more likely related to magnetic reconnection occurring in the lower-atmosphere part of the U-shape structures, whereas the reconnection in our observations appears to occur in a relatively higher part (possibly the lower corona) of the U-shape structures.
We noticed that a similar process has been theoretically and numerically investigated by \citet{2018ApJ...862L..24P} and \citet{2020ApJ...891...52S}, respectively. 

The formation process of coronal loops in our observations appears to be different from that in the numerical simulations of \citet{2014A&A...564A..12C}.
In this 3D MHD model, magnetic flux tubes expand into the corona during flux emergence. They found that the formation of million-degree coronal loops is triggered by an increase of the upward-directed Poynting flux at the loop footpoints due to the advection of the magnetic field in the photosphere. Ohmic heating associated with the coronal currents induced by the braided magnetic field lines leads to the formation of the high-temperature loops. In this model,  the magnetic flux tubes already exist in the corona before the formation of the million-degree coronal loops. On the contrary, in our observations the magnetic reconnection produces overlying field lines and heated plasma attached to these field lines, which then rise up to higher layers and appear as newly formed coronal loops. Such a process appears to be similar to the scenario proposed by \citet{2010A&A...510A..40H}.

\section{Summary}
\label{sec:sum}
From EUV observations of SDO/AIA and \ha\, line core observations of NVST, we have identified direct observational evidence for the formation of coronal loops through magnetic reconnection as new magnetic fluxes rise into the upper solar atmosphere. The reconnection occurs within a plasma sheet at the interface of two approaching loop-like structures. Converging motions of opposite-polarity magnetic fluxes and the subsequent flux cancellation are clearly observed from photospheric magnetograms. The reconnection results in the formation of overlying loops with typical coronal temperatures and low-lying small loops/fibrils above and below the plasma sheet, respectively. In the meantime, the transverse magnetic field in the photosphere is enhanced. After reconnection, the transverse field becomes weaker and the fibrils disappear, indicating the submergence of the low-lying loops. 

The EUV observations have revealed the presence of numerous bright plasma blobs in the plasma sheet. These blobs have an average width of 1.37 Mm, and they appear intermittently in the plasma sheet and move upward with projected velocities of $\sim$114\,\kms. Through a DEM analysis, the temperature, emission measure and density of the blobs have been found to be about 3\,MK, 2.0$\times$10$^{28}$\,cm$^{-5}$ and 1.2$\times$10$^{10}$\,cm$^{-3}$, respectively. We have performed a power spectral analysis for these blobs, and found a spectral index that is distinctly different from the expected one in a turbulent reconnection scenario.  In addition, we have also found plasma flows with speeds of 20 to 50\,\kms towards the footpoints of the newly formed coronal loops.

\acknowledgments

This work was supported by NSFC grants 11825301, 11790304, 11803002, 41874200, the Strategic Priority Research Program of CAS (grant XDA17040507), and grant 1916321TS00103201. H.C.C. was supported by the National Postdoctoral Program for Innovative Talents (BX20200013) and China Postdoctoral Science Foundation (2020M680201).
AIA and HMI are instruments onboard the Solar Dynamics Observatory, a mission for NASA's Living With a Star program.
The \ha\, data used in this paper were obtained with the New Vacuum Solar Telescope in Fuxian Solar Observatory of Yunnan Astronomical Observatories, CAS.
We thank Dr. Yang Guo for the fruitful discussion. 

\bibliographystyle{aasjournal}
\bibliography{bibliography}

\begin{thebibliography}{}
\expandafter\ifx\csname natexlab\endcsname\relax\def\natexlab#1{#1}\fi
\providecommand{\url}[1]{\href{#1}{#1}}
\providecommand{\dodoi}[1]{doi:~\href{http://doi.org/#1}{\nolinkurl{#1}}}
\providecommand{\doeprint}[1]{\href{http://ascl.net/#1}{\nolinkurl{http://ascl.net/#1}}}
\providecommand{\doarXiv}[1]{\href{https://arxiv.org/abs/#1}{\nolinkurl{https://arxiv.org/abs/#1}}}

\bibitem[{{B{\'a}rta} {et~al.}(2011){B{\'a}rta}, {B{\"u}chner}, {Karlick{\'y}},
  \& {Sk{\'a}la}}]{2011ApJ...737...24B}
{B{\'a}rta}, M., {B{\"u}chner}, J., {Karlick{\'y}}, M., \& {Sk{\'a}la}, J.
  2011, \apj, 737, 24, \dodoi{10.1088/0004-637X/737/1/24}

\bibitem[{{Borrero} {et~al.}(2011){Borrero}, {Tomczyk}, {Kubo},
  {Socas-Navarro}, {Schou}, {Couvidat}, \& {Bogart}}]{2011SoPh..273..267B}
{Borrero}, J.~M., {Tomczyk}, S., {Kubo}, M., {et~al.} 2011, \solphys, 273, 267,
  \dodoi{10.1007/s11207-010-9515-6}

\bibitem[{{Chen} {et~al.}(2014){Chen}, {Peter}, {Bingert}, \&
  {Cheung}}]{2014A&A...564A..12C}
{Chen}, F., {Peter}, H., {Bingert}, S., \& {Cheung}, M.~C.~M. 2014, \aap, 564,
  A12, \dodoi{10.1051/0004-6361/201322859}

\bibitem[{{Chen} {et~al.}(2019{\natexlab{a}}){Chen}, {Yang}, {Duan}, \&
  {Ji}}]{2019ApJ...879...74C}
{Chen}, H., {Yang}, J., {Duan}, Y., \& {Ji}, K. 2019{\natexlab{a}}, \apj, 879,
  74, \dodoi{10.3847/1538-4357/ab24ce}

\bibitem[{{Chen} {et~al.}(2019{\natexlab{b}}){Chen}, {Tian}, {Zhu}, {Samanta},
  {Wang}, \& {He}}]{2019ScChE..62.1555C}
{Chen}, Y., {Tian}, H., {Zhu}, X., {et~al.} 2019{\natexlab{b}}, Science China
  Technological Sciences, 62, 1555, \dodoi{10.1007/s11431-018-9471-6}

\bibitem[{{Chen} {et~al.}(2019{\natexlab{c}}){Chen}, {Tian}, {Peter},
  {Samanta}, {Yurchyshyn}, {Wang}, {Cao}, {Wang}, \&
  {He}}]{2019ApJ...875L..30C}
{Chen}, Y., {Tian}, H., {Peter}, H., {et~al.} 2019{\natexlab{c}}, \apjl, 875,
  L30, \dodoi{10.3847/2041-8213/ab18a4}

\bibitem[{{Cheng} {et~al.}(2018){Cheng}, {Li}, {Wan}, {Ding}, {Chen}, {Zhang},
  \& {Liu}}]{2018ApJ...866...64C}
{Cheng}, X., {Li}, Y., {Wan}, L.~F., {et~al.} 2018, \apj, 866, 64,
  \dodoi{10.3847/1538-4357/aadd16}

\bibitem[{{Cheung} {et~al.}(2015){Cheung}, {Boerner}, {Schrijver}, {Testa},
  {Chen}, {Peter}, \& {Malanushenko}}]{2015ApJ...807..143C}
{Cheung}, M. C.~M., {Boerner}, P., {Schrijver}, C.~J., {et~al.} 2015, \apj,
  807, 143, \dodoi{10.1088/0004-637X/807/2/143}

\bibitem[{{Chitta} {et~al.}(2020){Chitta}, {Peter}, {Priest}, \&
  {Solanki}}]{2020A&A...644A.130C}
{Chitta}, L.~P., {Peter}, H., {Priest}, E.~R., \& {Solanki}, S.~K. 2020, \aap,
  644, A130, \dodoi{10.1051/0004-6361/202039099}

\bibitem[{{Chitta} {et~al.}(2018){Chitta}, {Peter}, \&
  {Solanki}}]{2018A&A...615L...9C}
{Chitta}, L.~P., {Peter}, H., \& {Solanki}, S.~K. 2018, \aap, 615, L9,
  \dodoi{10.1051/0004-6361/201833404}

\bibitem[{{De Pontieu} {et~al.}(2014){De Pontieu}, {Title}, {Lemen}, {Kushner},
  {Akin}, {Allard}, {Berger}, {Boerner}, {Cheung}, {Chou}, {Drake}, {Duncan},
  {Freeland}, {Heyman}, {Hoffman}, {Hurlburt}, {Lindgren}, {Mathur}, {Rehse},
  {Sabolish}, {Seguin}, {Schrijver}, {Tarbell}, {W{\"u}lser}, {Wolfson},
  {Yanari}, {Mudge}, {Nguyen-Phuc}, {Timmons}, {van Bezooijen}, {Weingrod},
  {Brookner}, {Butcher}, {Dougherty}, {Eder}, {Knagenhjelm}, {Larsen},
  {Mansir}, {Phan}, {Boyle}, {Cheimets}, {DeLuca}, {Golub}, {Gates}, {Hertz},
  {McKillop}, {Park}, {Perry}, {Podgorski}, {Reeves}, {Saar}, {Testa}, {Tian},
  {Weber}, {Dunn}, {Eccles}, {Jaeggli}, {Kankelborg}, {Mashburn}, {Pust},
  {Springer}, {Carvalho}, {Kleint}, {Marmie}, {Mazmanian}, {Pereira}, {Sawyer},
  {Strong}, {Worden}, {Carlsson}, {Hansteen}, {Leenaarts}, {Wiesmann},
  {Aloise}, {Chu}, {Bush}, {Scherrer}, {Brekke}, {Martinez-Sykora}, {Lites},
  {McIntosh}, {Uitenbroek}, {Okamoto}, {Gummin}, {Auker}, {Jerram}, {Pool}, \&
  {Waltham}}]{2014SoPh..289.2733D}
{De Pontieu}, B., {Title}, A.~M., {Lemen}, J.~R., {et~al.} 2014, \solphys, 289,
  2733, \dodoi{10.1007/s11207-014-0485-y}

\bibitem[{{Ellerman}(1917)}]{1917ApJ....46..298E}
{Ellerman}, F. 1917, \apj, 46, 298, \dodoi{10.1086/142366}

\bibitem[{{Fan} {et~al.}(2018){Fan}, {He}, {Yan}, {Tomczyk}, {Tian}, {Song},
  {Wang}, \& {Zhang}}]{2018SoPh..293....6F}
{Fan}, S., {He}, J., {Yan}, L., {et~al.} 2018, \solphys, 293, 6,
  \dodoi{10.1007/s11207-017-1221-1}

\bibitem[{{Gou} {et~al.}(2019){Gou}, {Liu}, {Kliem}, {Wang}, \&
  {Veronig}}]{2019SciA....5.7004G}
{Gou}, T., {Liu}, R., {Kliem}, B., {Wang}, Y., \& {Veronig}, A.~M. 2019,
  Science Advances, 5, 7004, \dodoi{10.1126/sciadv.aau7004}

\bibitem[{{Grubecka} {et~al.}(2016){Grubecka}, {Schmieder}, {Berlicki},
  {Heinzel}, {Dalmasse}, \& {Mein}}]{2016A&A...593A..32G}
{Grubecka}, M., {Schmieder}, B., {Berlicki}, A., {et~al.} 2016, \aap, 593, A32,
  \dodoi{10.1051/0004-6361/201527358}

\bibitem[{{He} {et~al.}(2010){He}, {Marsch}, {Tu}, {Tian}, \&
  {Guo}}]{2010A&A...510A..40H}
{He}, J.~S., {Marsch}, E., {Tu}, C.~Y., {Tian}, H., \& {Guo}, L.~J. 2010, \aap,
  510, A40, \dodoi{10.1051/0004-6361/200913059}

\bibitem[{{Hoeksema} {et~al.}(2014){Hoeksema}, {Liu}, {Hayashi}, {Sun},
  {Schou}, {Couvidat}, {Norton}, {Bobra}, {Centeno}, {Leka}, {Barnes}, \&
  {Turmon}}]{2014SoPh..289.3483H}
{Hoeksema}, J.~T., {Liu}, Y., {Hayashi}, K., {et~al.} 2014, \solphys, 289,
  3483, \dodoi{10.1007/s11207-014-0516-8}

\bibitem[{{Hou} {et~al.}(2021){Hou}, {He}, {Zhu}, \&
  {Wang}}]{2021ApJ...908..237H}
{Hou}, C., {He}, J., {Zhu}, X., \& {Wang}, Y. 2021, \apj, 908, 237,
  \dodoi{10.3847/1538-4357/abd6f3}

\bibitem[{{Huang} {et~al.}(2018){Huang}, {Mou}, {Fu}, {Deng}, {Li}, \&
  {Xia}}]{2018ApJ...853L..26H}
{Huang}, Z., {Mou}, C., {Fu}, H., {et~al.} 2018, \apjl, 853, L26,
  \dodoi{10.3847/2041-8213/aaa88c}

\bibitem[{{Huang} {et~al.}(2015){Huang}, {Xia}, {Li}, \&
  {Madjarska}}]{2015ApJ...810...46H}
{Huang}, Z., {Xia}, L., {Li}, B., \& {Madjarska}, M.~S. 2015, \apj, 810, 46,
  \dodoi{10.1088/0004-637X/810/1/46}

\bibitem[{{Kazachenko} \& {Hudson}(2020)}]{2020ApJ...901...64K}
{Kazachenko}, M.~D., \& {Hudson}, H.~S. 2020, \apj, 901, 64,
  \dodoi{10.3847/1538-4357/abada6}

\bibitem[{{Leka} {et~al.}(2009){Leka}, {Barnes}, {Crouch}, {Metcalf}, {Gary},
  {Jing}, \& {Liu}}]{2009SoPh..260...83L}
{Leka}, K.~D., {Barnes}, G., {Crouch}, A.~D., {et~al.} 2009, \solphys, 260, 83,
  \dodoi{10.1007/s11207-009-9440-8}

\bibitem[{{Lemen} {et~al.}(2012){Lemen}, {Title}, {Akin}, {Boerner}, {Chou},
  {Drake}, {Duncan}, {Edwards}, {Friedlaender}, {Heyman}, {Hurlburt}, {Katz},
  {Kushner}, {Levay}, {Lindgren}, {Mathur}, {McFeaters}, {Mitchell}, {Rehse},
  {Schrijver}, {Springer}, {Stern}, {Tarbell}, {Wuelser}, {Wolfson}, {Yanari},
  {Bookbinder}, {Cheimets}, {Caldwell}, {Deluca}, {Gates}, {Golub}, {Park},
  {Podgorski}, {Bush}, {Scherrer}, {Gummin}, {Smith}, {Auker}, {Jerram},
  {Pool}, {Soufli}, {Windt}, {Beardsley}, {Clapp}, {Lang}, \&
  {Waltham}}]{2012SoPh..275...17L}
{Lemen}, J.~R., {Title}, A.~M., {Akin}, D.~J., {et~al.} 2012, \solphys, 275,
  17, \dodoi{10.1007/s11207-011-9776-8}

\bibitem[{{Lenz} {et~al.}(1999){Lenz}, {DeLuca}, {Golub}, {Rosner}, \&
  {Bookbinder}}]{1999ApJ...517L.155L}
{Lenz}, D.~D., {DeLuca}, E.~E., {Golub}, L., {Rosner}, R., \& {Bookbinder},
  J.~A. 1999, \apjl, 517, L155, \dodoi{10.1086/312045}

\bibitem[{{Li} {et~al.}(2016){Li}, {Zhang}, {Peter}, {Priest}, {Chen}, {Guo},
  {Chen}, \& {Mackay}}]{2016NatPh..12..847L}
{Li}, L., {Zhang}, J., {Peter}, H., {et~al.} 2016, Nature Physics, 12, 847,
  \dodoi{10.1038/nphys3768}

\bibitem[{{Li} {et~al.}(2018){Li}, {Xue}, {Ding}, {Cheng}, {Su}, {Feng},
  {Hong}, {Li}, \& {Gan}}]{2018ApJ...853L..15L}
{Li}, Y., {Xue}, J.~C., {Ding}, M.~D., {et~al.} 2018, \apjl, 853, L15,
  \dodoi{10.3847/2041-8213/aaa6c0}

\bibitem[{{Lin} {et~al.}(2009){Lin}, {Li}, {Ko}, \& {Raymond}}]{Lin2009}
{Lin}, J., {Li}, J., {Ko}, Y.~K., \& {Raymond}, J.~C. 2009, \apj, 693, 1666,
  \dodoi{10.1088/0004-637X/693/2/1666}

\bibitem[{Liu {et~al.}(2014)Liu, Xu, Gu, Wang, You, Shen, Lu, Jin, Chen, Lou,
  Li, Liu, Xu, Rao, Hu, Li, Fu, Wang, Bao, Wu, \& Zhang}]{Liu_2014}
Liu, Z., Xu, J., Gu, B.-Z., {et~al.} 2014, RAA, 14, 705,
  \dodoi{10.1088/1674-4527/14/6/009}

\bibitem[{{Metcalf}(1994)}]{1994SoPh..155..235M}
{Metcalf}, T.~R. 1994, \solphys, 155, 235, \dodoi{10.1007/BF00680593}

\bibitem[{{Mishra} {et~al.}(2020){Mishra}, {Srivastava}, \&
  {Chen}}]{2020SoPh..295..167M}
{Mishra}, S.~K., {Srivastava}, A.~K., \& {Chen}, P.~F. 2020, \solphys, 295,
  167, \dodoi{10.1007/s11207-020-01733-w}

\bibitem[{{Mok} {et~al.}(2008){Mok}, {Miki{\'c}}, {Lionello}, \&
  {Linker}}]{2008ApJ...679L.161M}
{Mok}, Y., {Miki{\'c}}, Z., {Lionello}, R., \& {Linker}, J.~A. 2008, \apjl,
  679, L161, \dodoi{10.1086/589440}

\bibitem[{{Nelson} {et~al.}(2016){Nelson}, {Doyle}, \&
  {Erd{\'e}lyi}}]{2016MNRAS.463.2190N}
{Nelson}, C.~J., {Doyle}, J.~G., \& {Erd{\'e}lyi}, R. 2016, \mnras, 463, 2190,
  \dodoi{10.1093/mnras/stw2034}

\bibitem[{{Ni} {et~al.}(2021){Ni}, {Chen}, {Peter}, {Tian}, \&
  {Lin}}]{2021A&A...646A..88N}
{Ni}, L., {Chen}, Y., {Peter}, H., {Tian}, H., \& {Lin}, J. 2021, \aap, 646,
  A88, \dodoi{10.1051/0004-6361/202039239}

\bibitem[{{Ni} {et~al.}(2020){Ni}, {Ji}, {Murphy}, \&
  {Jara-Almonte}}]{2020RSPSA.47690867N}
{Ni}, L., {Ji}, H., {Murphy}, N.~A., \& {Jara-Almonte}, J. 2020, Proceedings of
  the Royal Society of London Series A, 476, 90867,
  \dodoi{10.1098/rspa.2019.0867}

\bibitem[{{Ni} {et~al.}(2015){Ni}, {Kliem}, {Lin}, \&
  {Wu}}]{2015ApJ...799...79N}
{Ni}, L., {Kliem}, B., {Lin}, J., \& {Wu}, N. 2015, \apj, 799, 79,
  \dodoi{10.1088/0004-637X/799/1/79}

\bibitem[{{Pariat} {et~al.}(2004){Pariat}, {Aulanier}, {Schmieder},
  {Georgoulis}, {Rust}, \& {Bernasconi}}]{2004ApJ...614.1099P}
{Pariat}, E., {Aulanier}, G., {Schmieder}, B., {et~al.} 2004, \apj, 614, 1099,
  \dodoi{10.1086/423891}

\bibitem[{{Pesnell} {et~al.}(2012){Pesnell}, {Thompson}, \&
  {Chamberlin}}]{2012SoPh..275....3P}
{Pesnell}, W.~D., {Thompson}, B.~J., \& {Chamberlin}, P.~C. 2012, \solphys,
  275, 3, \dodoi{10.1007/s11207-011-9841-3}

\bibitem[{{Peter} {et~al.}(2019){Peter}, {Huang}, {Chitta}, \&
  {Young}}]{2019A&A...628A...8P}
{Peter}, H., {Huang}, Y.~M., {Chitta}, L.~P., \& {Young}, P.~R. 2019, \aap,
  628, A8, \dodoi{10.1051/0004-6361/201935820}

\bibitem[{{Peter} {et~al.}(2014){Peter}, {Tian}, {Curdt}, {Schmit}, {Innes},
  {De Pontieu}, {Lemen}, {Title}, {Boerner}, {Hurlburt}, {Tarbell}, {Wuelser},
  {Mart{\'\i}nez-Sykora}, {Kleint}, {Golub}, {McKillop}, {Reeves}, {Saar},
  {Testa}, {Kankelborg}, {Jaeggli}, {Carlsson}, \&
  {Hansteen}}]{2014Sci...346C.315P}
{Peter}, H., {Tian}, H., {Curdt}, W., {et~al.} 2014, Science, 346, 1255726,
  \dodoi{10.1126/science.1255726}

\bibitem[{{Priest} {et~al.}(2018){Priest}, {Chitta}, \&
  {Syntelis}}]{2018ApJ...862L..24P}
{Priest}, E.~R., {Chitta}, L.~P., \& {Syntelis}, P. 2018, \apjl, 862, L24,
  \dodoi{10.3847/2041-8213/aad4fc}

\bibitem[{{Reale}(2014)}]{2014LRSP...11....4R}
{Reale}, F. 2014, Living Reviews in Solar Physics, 11, 4,
  \dodoi{10.12942/lrsp-2014-4}

\bibitem[{{Samanta} {et~al.}(2021){Samanta}, {Tian}, {Chen}, {Reeves},
  {Cheung}, {Vourlidas}, \& {Banerjee}}]{2021Innov...200083S}
{Samanta}, T., {Tian}, H., {Chen}, B., {et~al.} 2021, The Innovation, 2,
  100083, \dodoi{10.1016/j.xinn.2021.100083}

\bibitem[{{Schou} {et~al.}(2012){Schou}, {Scherrer}, {Bush}, {Wachter},
  {Couvidat}, {Rabello-Soares}, {Bogart}, {Hoeksema}, {Liu}, {Duvall}, {Akin},
  {Allard}, {Miles}, {Rairden}, {Shine}, {Tarbell}, {Title}, {Wolfson},
  {Elmore}, {Norton}, \& {Tomczyk}}]{Schou2012hmi}
{Schou}, J., {Scherrer}, P.~H., {Bush}, R.~I., {et~al.} 2012, \solphys, 275,
  229, \dodoi{10.1007/s11207-011-9842-2}

\bibitem[{{Shen} {et~al.}(2011){Shen}, {Lin}, \&
  {Murphy}}]{2011ApJ...737...14S}
{Shen}, C., {Lin}, J., \& {Murphy}, N.~A. 2011, \apj, 737, 14,
  \dodoi{10.1088/0004-637X/737/1/14}

\bibitem[{{Song} {et~al.}(2020){Song}, {Tian}, {Zhu}, {Chen}, {Zhang}, \&
  {Zhang}}]{2020ApJ...893L..13S}
{Song}, Y., {Tian}, H., {Zhu}, X., {et~al.} 2020, \apjl, 893, L13,
  \dodoi{10.3847/2041-8213/ab83fa}

\bibitem[{{Su} {et~al.}(2018){Su}, {Veronig}, {Hannah}, {Cheung}, {Dennis},
  {Holman}, {Gan}, \& {Li}}]{2018ApJ...856L..17S}
{Su}, Y., {Veronig}, A.~M., {Hannah}, I.~G., {et~al.} 2018, \apjl, 856, L17,
  \dodoi{10.3847/2041-8213/aab436}

\bibitem[{{Su} {et~al.}(2013){Su}, {Veronig}, {Holman}, {Dennis}, {Wang},
  {Temmer}, \& {Gan}}]{2013NatPh...9..489S}
{Su}, Y., {Veronig}, A.~M., {Holman}, G.~D., {et~al.} 2013, Nature Physics, 9,
  489, \dodoi{10.1038/nphys2675}

\bibitem[{{Sun} {et~al.}(2015){Sun}, {Cheng}, {Ding}, {Guo}, {Priest},
  {Parnell}, {Edwards}, {Zhang}, {Chen}, \& {Fang}}]{2015NatCo...6.7598S}
{Sun}, J.~Q., {Cheng}, X., {Ding}, M.~D., {et~al.} 2015, Nature Communications,
  6, 7598, \dodoi{10.1038/ncomms8598}

\bibitem[{{Syntelis} \& {Priest}(2020)}]{2020ApJ...891...52S}
{Syntelis}, P., \& {Priest}, E.~R. 2020, \apj, 891, 52,
  \dodoi{10.3847/1538-4357/ab6ffc}

\bibitem[{{Takasao} {et~al.}(2012){Takasao}, {Asai}, {Isobe}, \&
  {Shibata}}]{2012ApJ...745L...6T}
{Takasao}, S., {Asai}, A., {Isobe}, H., \& {Shibata}, K. 2012, \apjl, 745, L6,
  \dodoi{10.1088/2041-8205/745/1/L6}

\bibitem[{{Takasao} {et~al.}(2016){Takasao}, {Asai}, {Isobe}, \&
  {Shibata}}]{2016ApJ...828..103T}
---. 2016, \apj, 828, 103, \dodoi{10.3847/0004-637X/828/2/103}

\bibitem[{{Tian} {et~al.}(2014){Tian}, {Li}, {Reeves}, {Raymond}, {Guo}, {Liu},
  {Chen}, \& {Murphy}}]{2014ApJ...797L..14T}
{Tian}, H., {Li}, G., {Reeves}, K.~K., {et~al.} 2014, \apjl, 797, L14,
  \dodoi{10.1088/2041-8205/797/2/L14}

\bibitem[{{Tian} {et~al.}(2016){Tian}, {Xu}, {He}, \&
  {Madsen}}]{2016ApJ...824...96T}
{Tian}, H., {Xu}, Z., {He}, J., \& {Madsen}, C. 2016, \apj, 824, 96,
  \dodoi{10.3847/0004-637X/824/2/96}

\bibitem[{{Tian} {et~al.}(2018{\natexlab{a}}){Tian}, {Zhu}, {Peter}, {Zhao},
  {Samanta}, \& {Chen}}]{2018ApJ...854..174T}
{Tian}, H., {Zhu}, X., {Peter}, H., {et~al.} 2018{\natexlab{a}}, \apj, 854,
  174, \dodoi{10.3847/1538-4357/aaaae6}

\bibitem[{{Tian} {et~al.}(2018{\natexlab{b}}){Tian}, {Yurchyshyn}, {Peter},
  {Solanki}, {Young}, {Ni}, {Cao}, {Ji}, {Zhu}, {Zhang}, {Samanta}, {Song},
  {He}, {Wang}, \& {Chen}}]{2018ApJ...854...92T}
{Tian}, H., {Yurchyshyn}, V., {Peter}, H., {et~al.} 2018{\natexlab{b}}, \apj,
  854, 92, \dodoi{10.3847/1538-4357/aaa89d}

\bibitem[{{Tripathi}(2021)}]{2021ApJ...909..105T}
{Tripathi}, D. 2021, \apj, 909, 105, \dodoi{10.3847/1538-4357/abdd2e}

\bibitem[{{Verscharen} {et~al.}(2019){Verscharen}, {Klein}, \&
  {Maruca}}]{2019LRSP...16....5V}
{Verscharen}, D., {Klein}, K.~G., \& {Maruca}, B.~A. 2019, Living Reviews in
  Solar Physics, 16, 5, \dodoi{10.1007/s41116-019-0021-0}

\bibitem[{{Vissers} {et~al.}(2015){Vissers}, {Rouppe van der Voort}, {Rutten},
  {Carlsson}, \& {De Pontieu}}]{2015ApJ...812...11V}
{Vissers}, G.~J.~M., {Rouppe van der Voort}, L.~H.~M., {Rutten}, R.~J.,
  {Carlsson}, M., \& {De Pontieu}, B. 2015, \apj, 812, 11,
  \dodoi{10.1088/0004-637X/812/1/11}

\bibitem[{{Wang} \& {Shi}(1993)}]{1993SoPh..143..119W}
{Wang}, J., \& {Shi}, Z. 1993, \solphys, 143, 119, \dodoi{10.1007/BF00619100}

\bibitem[{{Winebarger} {et~al.}(2011){Winebarger}, {Schmelz}, {Warren}, {Saar},
  \& {Kashyap}}]{Winebarger2011}
{Winebarger}, A.~R., {Schmelz}, J.~T., {Warren}, H.~P., {Saar}, S.~H., \&
  {Kashyap}, V.~L. 2011, \apj, 740, 2, \dodoi{10.1088/0004-637X/740/1/2}

\bibitem[{{Xiang} {et~al.}(2016){Xiang}, {Liu}, \& {Jin}}]{2016NewA...49....8X}
{Xiang}, Y.-y., {Liu}, Z., \& {Jin}, Z.-y. 2016, \na, 49, 8,
  \dodoi{10.1016/j.newast.2016.05.002}

\bibitem[{{Xie} {et~al.}(2017){Xie}, {Madjarska}, {Li}, {Huang}, {Xia},
  {Wiegelmann}, {Fu}, \& {Mou}}]{Xie2017}
{Xie}, H., {Madjarska}, M.~S., {Li}, B., {et~al.} 2017, \apj, 842, 38,
  \dodoi{10.3847/1538-4357/aa7415}

\bibitem[{{Xue} {et~al.}(2020{\natexlab{a}}){Xue}, {Su}, {Li}, \&
  {Zhao}}]{2020ApJ...898...88X}
{Xue}, J., {Su}, Y., {Li}, H., \& {Zhao}, X. 2020{\natexlab{a}}, \apj, 898, 88,
  \dodoi{10.3847/1538-4357/ab9a3d}

\bibitem[{{Xue} {et~al.}(2018){Xue}, {Yan}, {Yang}, {Wang}, {Feng}, {Li}, {Ji},
  \& {Zhao}}]{2018ApJ...858L...4X}
{Xue}, Z., {Yan}, X., {Yang}, L., {et~al.} 2018, \apjl, 858, L4,
  \dodoi{10.3847/2041-8213/aabe77}

\bibitem[{{Xue} {et~al.}(2020{\natexlab{b}}){Xue}, {Yan}, {Yang}, {Wang}, {Li},
  \& {Zhao}}]{2020A&A...633A.121X}
---. 2020{\natexlab{b}}, \aap, 633, A121, \dodoi{10.1051/0004-6361/201936969}

\bibitem[{{Xue} {et~al.}(2016){Xue}, {Yan}, {Cheng}, {Yang}, {Su}, {Kliem},
  {Zhang}, {Liu}, {Bi}, {Xiang}, {Yang}, \& {Zhao}}]{2016NatCo...711837X}
{Xue}, Z., {Yan}, X., {Cheng}, X., {et~al.} 2016, Nature Communications, 7,
  11837, \dodoi{10.1038/ncomms11837}

\bibitem[{{Yan} {et~al.}(2015){Yan}, {He}, {Xia}, \&
  {Jiao}}]{2015ApJ...804...69Y}
{Yan}, L., {He}, J., {Xia}, L., \& {Jiao}, F. 2015, \apj, 804, 69,
  \dodoi{10.1088/0004-637X/804/1/69}

\bibitem[{{Yan} {et~al.}(2020){Yan}, {Liu}, {Zhang}, \& {Xu}}]{Yan2020}
{Yan}, X., {Liu}, Z., {Zhang}, J., \& {Xu}, Z. 2020, Science China
  Technological Sciences, 63, 1656, \dodoi{10.1007/s11431-019-1463-6}

\bibitem[{{Yan} {et~al.}(2018){Yan}, {Yang}, {Xue}, {Mei}, {Kong}, {Wang}, \&
  {Li}}]{2018ApJ...853L..18Y}
{Yan}, X.~L., {Yang}, L.~H., {Xue}, Z.~K., {et~al.} 2018, \apjl, 853, L18,
  \dodoi{10.3847/2041-8213/aaa6c2}

\bibitem[{{Yang} {et~al.}(2018){Yang}, {Yang}, {Bi}, {Hong}, {Li}, {Xu}, \&
  {Chen}}]{2018ApJ...861..135Y}
{Yang}, B., {Yang}, J., {Bi}, Y., {et~al.} 2018, \apj, 861, 135,
  \dodoi{10.3847/1538-4357/aac37f}

\bibitem[{{Yang} {et~al.}(2020){Yang}, {Li}, {Guo}, {Li}, {Li}, {He}, {Zhang},
  \& {Feng}}]{2020ApJ...901L..22Y}
{Yang}, L., {Li}, H., {Guo}, F., {et~al.} 2020, \apjl, 901, L22,
  \dodoi{10.3847/2041-8213/abb76b}

\bibitem[{{Yang} {et~al.}(2015){Yang}, {Zhang}, \&
  {Xiang}}]{2015ApJ...798L..11Y}
{Yang}, S., {Zhang}, J., \& {Xiang}, Y. 2015, \apjl, 798, L11,
  \dodoi{10.1088/2041-8205/798/1/L11}

\bibitem[{{Young} {et~al.}(2018){Young}, {Tian}, {Peter}, {Rutten}, {Nelson},
  {Huang}, {Schmieder}, {Vissers}, {Toriumi}, {Rouppe van der Voort},
  {Madjarska}, {Danilovic}, {Berlicki}, {Chitta}, {Cheung}, {Madsen},
  {Reardon}, {Katsukawa}, \& {Heinzel}}]{2018SSRv..214..120Y}
{Young}, P.~R., {Tian}, H., {Peter}, H., {et~al.} 2018, \ssr, 214, 120,
  \dodoi{10.1007/s11214-018-0551-0}

\bibitem[{{Zhang} \& {Ni}(2019)}]{2019ApJ...870..113Z}
{Zhang}, Q.~M., \& {Ni}, L. 2019, \apj, 870, 113,
  \dodoi{10.3847/1538-4357/aaf391}

\bibitem[{{Zhao} {et~al.}(2017){Zhao}, {Schmieder}, {Li}, {Pariat}, {Zhu},
  {Feng}, \& {Grubecka}}]{2017ApJ...836...52Z}
{Zhao}, J., {Schmieder}, B., {Li}, H., {et~al.} 2017, \apj, 836, 52,
  \dodoi{10.3847/1538-4357/836/1/52}

\bibitem[{{Zhu} \& {Wiegelmann}(2018)}]{2018ApJ...866..130Z}
{Zhu}, X., \& {Wiegelmann}, T. 2018, \apj, 866, 130,
  \dodoi{10.3847/1538-4357/aadf7f}

\bibitem[{{Zhu} \& {Wiegelmann}(2019)}]{2019A&A...631A.162Z}
---. 2019, \aap, 631, A162, \dodoi{10.1051/0004-6361/201936433}

\end{thebibliography}

\end{document}